\documentclass[conference]{IEEEtran}

\IEEEoverridecommandlockouts
\usepackage{cite}

\ifCLASSINFOpdf
   \usepackage[pdftex]{graphicx}
\else
   \usepackage[dvips]{graphicx}
   \DeclareGraphicsExtensions{.eps}
\fi
\usepackage{grffile}
\usepackage{epstopdf}

\usepackage[cmex10]{amsmath}

\usepackage{color}

\begin{document}
%
\title{Sliding Window-Based Contention Resolution Diversity Slotted ALOHA}


\author{\IEEEauthorblockN{Alessio Meloni\thanks{\copyright 2012 IEEE. The IEEE copyright notice applies.\ DOI: 10.1109/GLOCOM.2012.6503624} and Maurizio Murroni}
\IEEEauthorblockA{DIEE - Dept. of Electrical and Electronic Engineering\\
University of Cagliari\\
Piazza D'Armi, 09123 Cagliari, Italy\\
Email: \{alessio.meloni\}\{murroni\}@diee.unica.it}
\and
\IEEEauthorblockN{Christian Kissling and Matteo Berioli}
\IEEEauthorblockA{Department of Digital Networks\\
German Aerospace Center (DLR)\\
Oberpfaffenhofen, 82234 Germany\\
Email: \{Christian.Kissling\}\{matteo.berioli\}@dlr.de}

}


%


\maketitle
\begin{abstract}
Contention Resolution Diversity Slotted ALOHA (CRDSA) and its burst degree optimizations (CRDSA++, IRSA) make use of MAC burst repetitions and Interference Cancellation (IC) making possible to reach throughput values as high as $T \simeq 0.8$ in practical implementations, whereas for the traditional slotted ALOHA $T \simeq 0.37$. However, these new techniques introduce a frame-based access to the channel that limits the performance in terms of throughput and packet delivery delay. In this paper, a new technique named Sliding Window CRDSA (SW-CRDSA) and its counterpart for irregular repetitions (SW-IRSA) are introduced in order to exploit the advantages of MAC burst repetition and Interference Cancellation (IC) with an unframed access scheme. Numerical results are also provided in order to validate the statement of better performance.
\end{abstract}


%
\IEEEpeerreviewmaketitle
\section{Introduction}
In a multi-access channel for satellite communications with no pre-assigned resources, the possible ways to share the channel are basically two:
1) the channel is managed by a central entity that gives permission to users to transmit upon request, i.e. a reservation mechanism known as DAMA (Demand Assignment Multiple Access) is used;
2) no coordination among users exists, therefore each user sends its data according to a local algorithm (Random Access).
DAMA avoids the case of collisions among data sent from different users. However, reservation mechanisms need
time to be accomplished, since a three-way handshake or similar is needed. In other words, each user sends a capacity request and has to wait for a feedback in order to know whether it can transmit or not. This waiting time is not negligible and sometimes not acceptable when long propagation delays are present as in satellite communications, especially if the traffic from each user is bursty. In fact, in a typical satellite communication scenario with bent-pipe Geostationary Orbit (GEO) satellites, the minimum achievable value corresponding to a three-hop delay is $\approx 750\ ms$. Therefore DAMA is convenient only when single users have a medium or high amount of traffic to send. On the other hand, if transmissions from users are bursty as in the case of consumer type of interactive satellite terminals, a Random Access technique may be preferred, although the possibility of data collision is present.

Concerning slot-based random access techniques, Slotted Aloha (SA) \cite{SWCRDSA:RobertsALOHA} \cite{SWCRDSA:AbramsonALOHA} and Diversity Slotted Aloha (DSA) \cite{SWCRDSA:DiversityALOHA} are used indeed in satellite standards \cite{SWCRDSA:DVB-RCS} \cite{SWCRDSA:IPoS} whenever small amounts of data need to be sent. Defining $G$ as the normalized average number of packet transmissions per slot (namely \textit{MAC channel load}), SA reaches a peak throughput $T=1/e \simeq 0.37 \frac{pkt}{slot}$ for $G=1$, while DSA guarantees higher throughput up to moderate loads, thanks to diversity (each packet is sent more than once in different slots). Lately, a new technique exploiting the diversity of DSA by means of Interference Cancellation (IC), called Contention Resolution Diversity Slotted ALOHA (CRDSA) \cite{SWCRDSA:CRDSA1}, and some burst repetition optimizations named CRDSA++ \cite{SWCRDSA:CRDSA2} \cite{SWCRDSA:CRDSA3}, and IRSA \cite{SWCRDSA:IRSA1}  have been introduced \footnote{In the recent past, the concept of Contention Resolution has also been extended to the case of unslotted ALOHA (CRA) \cite{SWCRDSA:CRA} and to the case in which bursts are segmented and encoded prior to transmission (CSA) \cite{SWCRDSA:IRSAcoded}.}. 
These new random access schemes work as follows. Consider MAC frames composed of $N_f$ slots of duration $T_{slot}$. A finite number of users $N_{u}=GN_f$ attempts a packet transmission in a frame by sending a certain number of copies of the packet (\textit{instances}\footnote{In this paper, the terms packet and burst are used interchangeably. Same thing for the terms copy and instance.}) within the $N_f$ slots of that frame. Each instance contains a pointer to the slots where the other instances are, so that if at least one instance of a packet does not interefere with instances of the other packets and this instance is correctly received, the potential interference contribution of the other instances of the same packet can be removed from the other slots. This process is called Interference Cancellation\footnote{Additional details of the implementation of the IC mechanism are provided in \cite{SWCRDSA:CRDSA1} and \cite{SWCRDSA:IRSA1}.}(IC) and allows to restore the content of packets that initially had all their instances interfering. Moreover, the IC process can be iteratively repeated for the restored packets too, in order to successfully decode as many packets as possible. Therefore IC may allow the decoding of packets having all their instances interfering, yielding to better performance in terms of throughput. For this reason, these techniques are currently investigated within the Digital Video Broadcasting (DVB) for the Return Channel via Satellite (RCS) of the next generation of interactive satellite services \cite{NGrcs1}\cite{NGrcs2}.

However CRDSA, CRDSA++ and IRSA group slots in frames, implying that each user has to wait the beginning of the next frame to start sending its packets. This introduces an undesirable component of delay that is not present in SA and DSA, in which a packet (or the first instance of a packet) is typically sent in the next slot as soon as it is ready for transmission.
Also the throughput performance is limited by frames, because users transmitting in the same frame share the same set of eligible slots to place their instances and this increases in a way the probability of unsolvable collisions. Therefore the idea for this new technique arises from the need of an unframed Contention Resolution Diversity Slotted ALOHA technique capable of achieving better throughput and packet delivery delay than in the framed case.

The paper is organized as follows. In Section II the proposed RA scheme is presented. In Section III the advantages of the proposed technique are illustrated in an intuitive manner. Section IV motivates the degree distribution choices for the case of irregular repetitions. Section V deals with the adopted simulation approach. Finally, numerical results are compared in Section VI, in order to show performance improvements and to highlight the importance of proper selection for the key parameters of this technique. Section VII concludes the paper.

\section{Proposed random access scheme}

Consider a typical scenario such as the return link of a satellite access network, consisting of several user terminals communicating via satellite to the gateway located in a ground station. The proposed random access technique works in the following way: as soon as a user has a packet available for transmission, the first instance is sent in the next available slot while the other $l-1$ copies for the same packet are placed in the next $N_{sw}-1$ slots with equally distributed probability. $N_{sw}$ represents the number of successive slots (comprehensive of the one with the first instance) in which a certain user can place its packet instances and in this sense it can be considered as counterpart of the frame. We refer to this set of $N_{sw}$ slots as Sliding Window, to underline that time after time, depending on their arrival, the set of slots considered by users to place the instances of their packets is gradually sliding.
Therefore, only packets ready to be transmitted within the same slot interval have the same set, while those having an arrival difference of $ \lceil T_{ij}/T_{slot} \rceil$ slots ($T_{ij}$ indicates the time difference between packet \textit{i} and packet \textit{j} ) will share just $N_{sw} - \lceil T_{ij}/T_{slot} \rceil $ slots. At the receiver side, an IC iterative process similar to the one for FR is started at the end of each slot or, in a more general case, with a period that is multiple of $T_{slot}$.

\begin{figure}[t!]
\includegraphics [width=\columnwidth] {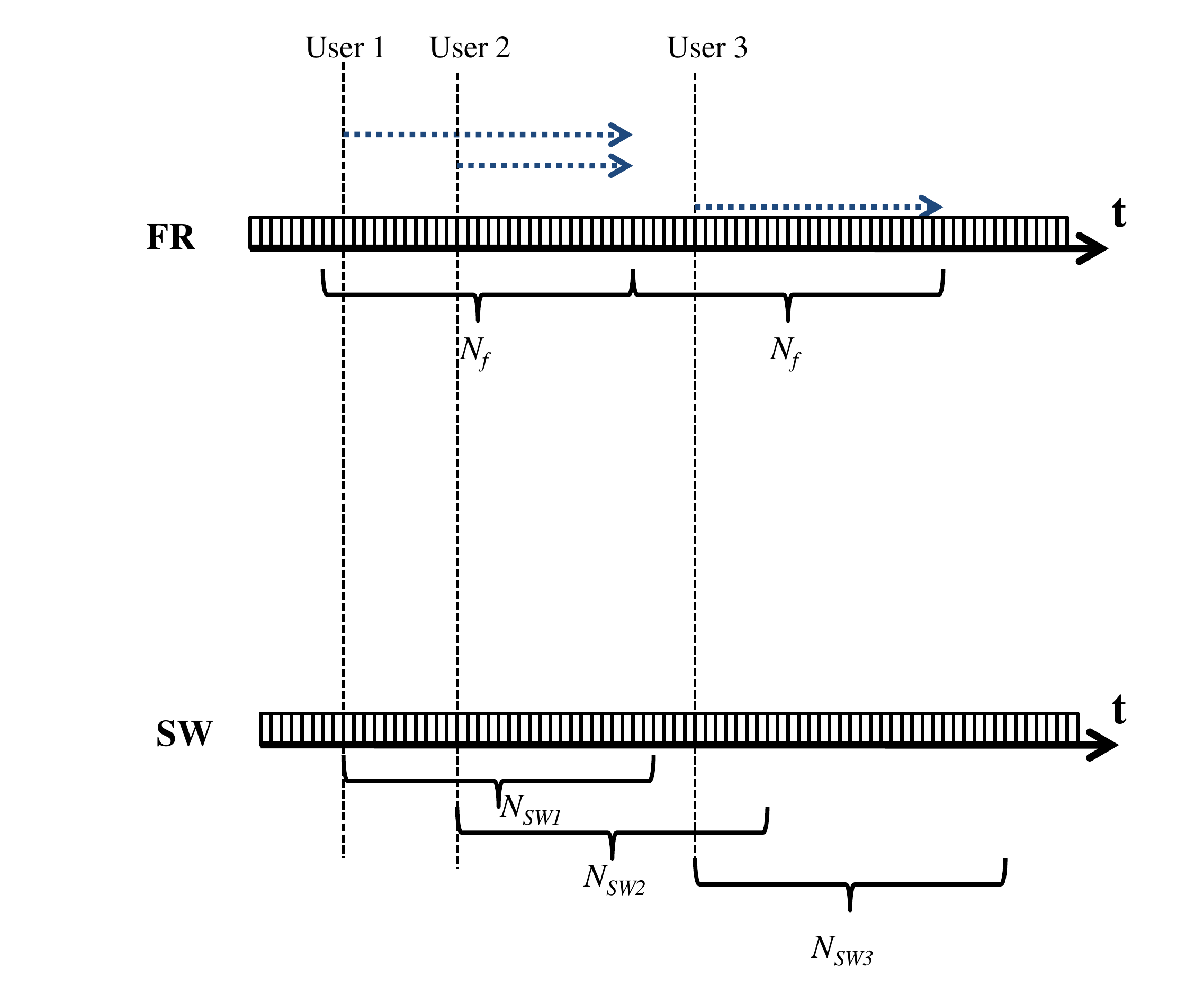}
\caption{Example of access to the channel for FR and SW. Dotted arrows indicate the waiting time for the beginning of the next frame in the FR case.}
\label{TXside}
\end{figure}

\begin{figure}[t!]
\includegraphics [width=\columnwidth] {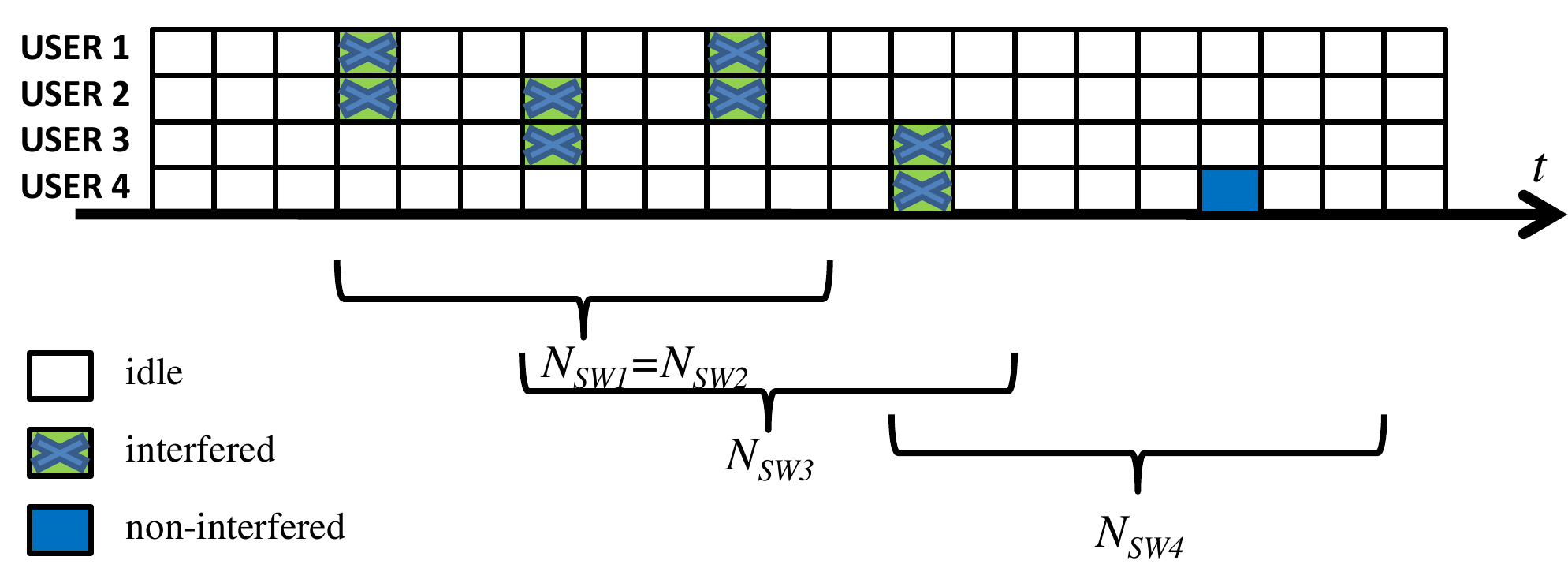}
\caption{Example of SW transmission using IRSA.}
\label{fig:Example}
\end{figure}

\section{Advantages of the proposed scheme}

\subsection{Throughput}

Figure~\ref{TXside} illustrates the difference between the techniques that group the slots in frames (from now on indicated as FR) and those using a Sliding Window (from now on indicated as SW).  In FR, packets ready to be transmitted would tipically wait until the beginning of the next frame to start sending their instances (the waiting interval is indicated with a dotted arrow). This implies that packets ready to be transmitted within the same frame, randomly choose the slots to place their instances from an identical set of slots (i.e. the slots of the frame), and a so called stopping set is created with a certain probability. A \textit{stopping set} \cite{SWCRDSA:SS} can be defined as a set of packets such that a certain set of slots contains all the instances of those packets and each slot contains at least two instances of different packets belonging to the stopping set.\\ 
On the contrary, as already described in the previous section and differently from FR, in the SW case users consider the set of the next $N_{sw}$ slots immediately after packet generation as set to put their packet instances. Therefore each of them has a different slots' set to put the instances of their packets, unless more than one packet was ready for transmission within the same slot interval. The result is that the probability to form a stopping set in SW depends also on the moment in which packets are ready for transmission; thus a lower probability to have a stopping set involving a certain set of users is found. In fact, if the equivalent graph representation given in \cite{SWCRDSA:IRSA1} is considered, the resulting graph for SW has infinite size and a more tree-like structure than the corresponding graph for FR, demonstrating that the probability of unsolvable collisions is smaller if the same settings (e.g. number of instances per packet) and $N_{sw}=N_f$ are assumed.
Consider Figure~\ref{fig:Example}, representing an example with irregular number of replicas. If perfect channel estimation and interference cancellation are assumed, each slot can be in one of three states:

\begin{itemize}
\item{no instances have been transmitted in a given slot, thus the slot is idle;}
\item{only 1 instance has been transmitted in a given slot, thus the packet did not interefere and is correctly decoded;}
\item{more than 1 instance has been transmitted in a given slot, thus interfering and resulting in loss of all the instances in that slot.}
\end{itemize}

User 1's packet instances are all interfering as well as User 2's and User 3's packet instances, but User 4 has an instance that did not collide. Therefore the content of the packet from User 3 can be restored thanks to IC of User 4's instances and in a waterfall manner also User 2's and User 1's packets will be restored. Notice that the IC process can be done only if all the slots containing the interfered instances are still memorized at the receiver. Thus, the receiver needs to memorize more than $N_{sw}$ slots so that restorable packet's instances are not lost. While adding a cost due to the need to keep in memory more packets at the receiver, this results in a better throughput performance as it will be shown hereinafter.

\subsection{Packet Delivery Delay}

In SW, the usage of a different Channel Access Algorithm allows users to send their packets without waiting for the beginning of the next frame, whereas in the FR case each user has to wait a time between 0 and $T_f$ (frame time duration) to start transmitting. 
Moreover, there is another delay component comprised between $T_{slot}$ and $T_f$, that depends on the placement of the packet instances over the frame, on the probability of recovering them at each IC process and on the frequency in the employment of the IC process (e.g. at the end of each frame or at the end of each slot). These two components together with the propagation delay $T_p$ represent the total packet delivery delay. In summary, the range for the packet delay $T_d$ in the FR case is
   
\begin{equation}
 T_p+T_{slot} < T_d \leq T_p + 2\ T_f
\end{equation}

In the proposed technique instead, the time delay varies from $T_p+T_{slot}$ to $T_p+T_{rx}$, where $T_{rx}$ is the number of slots memorized at the receiver ($N_{rx}$) times the slot duration $T_{slot}$. Even though this range is wider, it will be shown by means of simulations that the average delay in the SW case is always smaller than the one for FR, if $N_f=N_{sw}$ is assumed.

\section{Remarks on Irregular Repetitions for SW}

In FR, the use of irregular repetitions (IRSA) can yield to even better throughput results than using regular repetitions \cite{SWCRDSA:IRSA1}. Therefore, we want to extend the concept of IRSA also to the case of SW in order to demonstrate that SW overcomes FR both in delay and in throughput performance also when using irregular burst repetitions. Considering a certain maximum number of instances per packet, our choice has been to use some of the repetition distributions given in \cite{SWCRDSA:IRSA1}:

\begin{itemize}

\item $\Lambda (x)\ =\ 0.5102\ x^2\ +\ 0.4898\ x^4$ for maximum number of instances per packet equal to 4;

\item $\Lambda (x)\ =\ 0.5\ x^2\ +\ 0.28\ x^3\ +\ 0.22\ x^8$ for maximum number of instances per packet equal to 8;
 
\end{itemize}
where each term $\Psi_l x^l$ of the polynomial indicates that the probability of having $l$ burst instances (namely burst degree) for a certain packet is $\Psi_l$.\

The motivation driving to this choice is the intention to use the results obtained in \cite{SWCRDSA:IRSA1} for optimized burst repetitions, in order to have a comparison between SW and the best FR case for a given maximum number of instances per packet.
Moreover, it is expected that the repetition optimizations brought in \cite{SWCRDSA:IRSA1} for an infinite frame size are still valid for SW on a first approximation. In fact, in the asymptotic case of SW ($N_{sw}=\infty$) we can assume the entire history of the transmission equal to a frame of infinite size ($N_f=\infty$), that is also the asymptotic case for FR. Assuming that these distribution optimizations for FR calculated with an asymptotic setting remain valid also for a finite sliding window size (as done in \cite{SWCRDSA:IRSA1} for a finite frame size), the only thing that needs to be demonstrated is that the distribution of packet's instances over the slots in FR and SW are equal. In both cases, the polynomial representation of the degree distribution from the perspective of the slots is

\begin {equation}
\Psi(x) = \sum_{l} \Psi_l x^l 
\end {equation}

\begin {equation}
\Psi_l = \lim_{N_u\rightarrow \infty} \binom{N_u}{l}\bigg(\frac{\Psi'(1)}{N_u}\bigg)^l\bigg(1-\frac{\Psi'(1)}{N_u}\bigg)^{N_u-l}
\end {equation}

where $\frac{\Psi'(1)}{N_u} = P_{UinS}$ is the probability that a generic user puts an instance in a given slot.

Therefore we need to demonstrate that $P_{UinS}$ is equal for the two cases (i.e. $P_{UinS}^{FR} = P_{UinS}^{SW}$) . Let's define:

\begin{itemize}

\item $P_{i}^N = \frac{1}{N-i+1}$   probability of putting the $i^{th}$ instance in a given slot over $N$ possible slots (the $i^{th}$ instance can be put only in those slots not yet occupied by instances of the same packet);

\item $P_{NOT(j)}^N = \frac{N-1}{N} \frac{N-2}{N-1} \cdot\cdot\cdot \frac{N-j}{N-j+1} = \frac{N-j}{N}$   probability that none of j instances has been put in a given slot over $N$ possible slots;

\item $P_{first}^N = \frac{N{sw}-1}{N}$   probability that the first instance in the case of SW has been put no more than $N_{sw}-1$ slots before, so that one of the other instances can be put in the considered slot with a certain probability greater than zero (i.e. the considered slot belongs to the sliding window of that packet).

\end{itemize}

The probability that a generic user sends a burst copy within a given slot in the FR case is

\begin {equation}
\begin {split}
P_{UinS}^{FR} = P_{1}^{N_f} + P_{NOT1}^{N_f}P_{2}^{N_f} +\\+ ... + P_{NOT(l-1)}^{N_f} P_{l}^{N_f} = \frac{l}{N_f}
\end {split}
\end {equation}\\

while in the SW case

\begin {equation}
\begin {split}
P_{UinS}^{SW} = P_{1}^{N_f} + P_{first}^{N_{f}}(P_{NOT1}^{N_{sw}-1}P_{2}^{N_{sw}-1} +\\+ ... + P_{NOT(l-1)}^{N_{sw}-1} P_{l}^{N_{sw}-1}) = \frac{l}{N_f}
\end {split}
\end {equation}\\

Therefore $P_{UinS}^{FR}=P_{UinS}^{SW}$ that demonstrates the validity of the assumption.

\section{Simulation Approach}
In order to compare FR and SW, numerical simulations have been performed. All the simulations regard only the MAC layer. Moreover perfect channel estimation and interference cancellation have been assumed, so that bursts are either colliding or correctly
received. This means that capture effects\footnote{Capture effects may lead to better results especially in scenarios where power unbalance among different users is present. In fact, in this case a packet might be decodable even though it belongs to a stopping set.} and sources of disturbance such as noise have not been considered ($SNR=\infty$).
Differently from usual simulations for FR, we do not suppose a fixed channel load on each frame but  Poisson Arrivals with a given mean arrival rate $\bar{\lambda}$ (mean number of packet arrivals per slot interval). In fact, to achieve comparable results for FR and SW, the same assumptions on packet arrivals have to be made. Since in the case of SW it is not possible to fragment the entire history of the transmission into smaller independent parts (frames), a theoretically infinite time has to be considered together with a realistic packet arrivals distribution. For this reason, we assume an infinite population generating packets according to a Poisson distribution with a given mean $\bar{\lambda}$, that is the commonly used model for random access communications when packet arrivals are independent from each other and can be generated at any time with equal probability. Therefore, the number of users on each frame for FR will vary according to a Poisson distribution with mean $(\bar{\lambda}\cdot N_f)$.

Concerning the settings for our simulations, we consider a satellite communication system with GEO bent-pipe satellites with $T_{slot}=1\ ms$ assumed as time unit and propagation time from the source to the gateway $T_{p}=250\ ms$. Moreover, the receiver starts an IC process at the end of each slot both in the case of FR and in the case of SW, and waiting intervals for the beginning of the next slot are assumed to be negligible.
Concerning the maximum number of iterations for the IC iterative process, we have assumed $I_{max}=50$, a value big enough in order to make sure that the results are independent on the number of iterations of the IC process, since this is not the aim of our paper. Finally, the simulations have been performed for an \textit{open loop} communication scenario. Thus neither congestion control nor retransmissions have been considered.
 
\begin{figure}[ht!]
\centering
\includegraphics [width=0.9 \columnwidth] {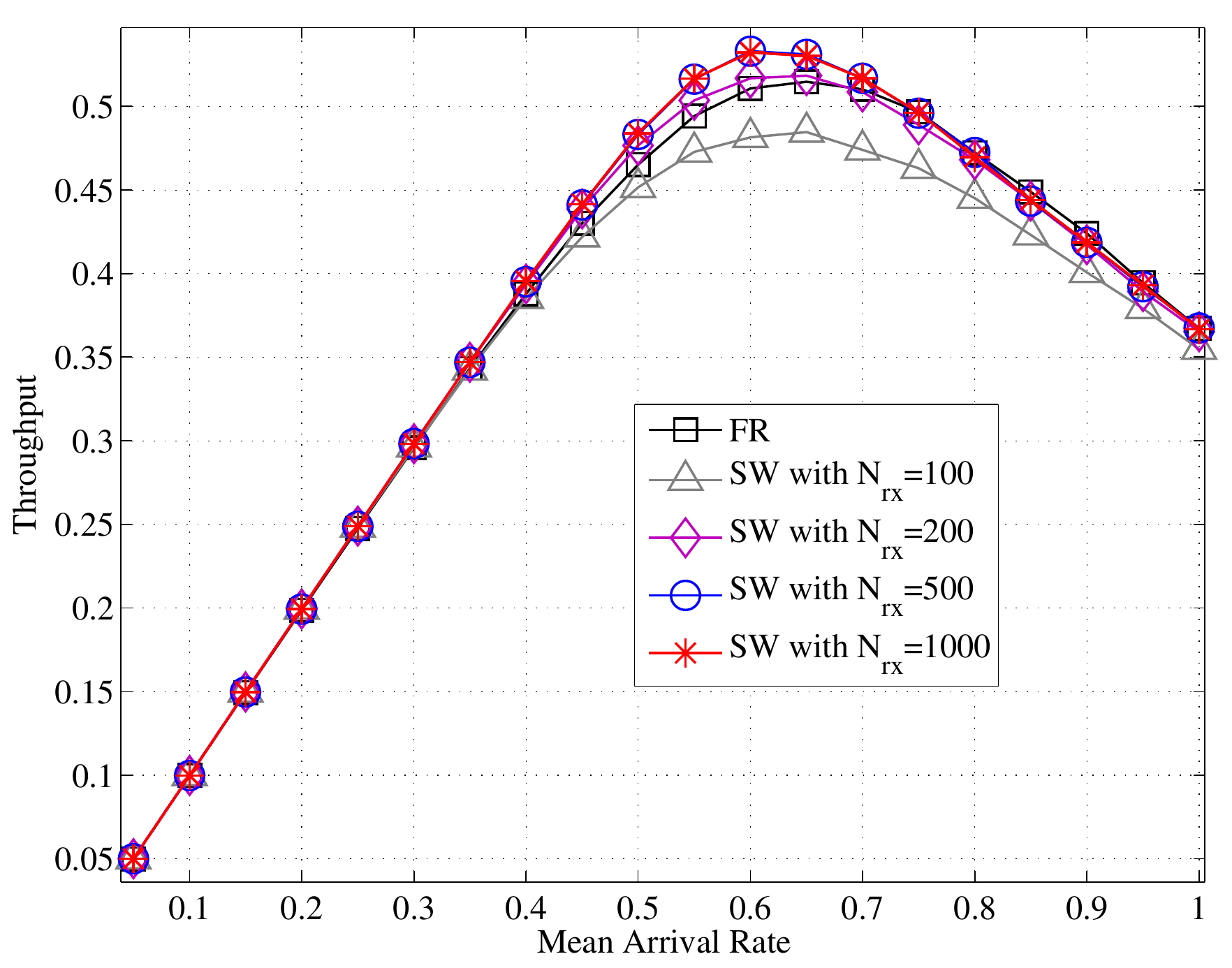}
\caption{Simulation results for the throughput in case of CRDSA (2 instances) with $N_f=N_{sw}=100$ slots.}
\label{incrNrx}
\end{figure}

\section{Numerical Results}

\subsection{Memory Size at the Receiver}
\label{memsize}
The first thing we want to highlight is the importance of having a memorization of more than $N_{sw}$ slots at the receiver. Consider Figure~\ref{incrNrx}, showing the throughput in the case of regular burst degree distribution with two instances per packet (CRDSA). For a memorization of slots at the receiver of the same size of the SW, the resulting throughput is worse than in FR. However, the performance with SW already overtakes the one in the framed case for $N_{rx} = 2N_{sw}$ and the throughput increases even more for bigger $N_{rx}$ values. This happens up to a point (in this case five times $N_{sw}$) in which there is no more improvement for the throughput even though bigger $N_{rx}$ values are assumed. This depends on the fact that the possibility of restoring packets is no longer dependent on the size of the receiver. In other words, even though an infinite number of slots is memorized at the receiver, those packets would not be decoded because the problem concerns the presence of stopping sets and is not related to the receiver size.

\subsection{Size of the Sliding Window}

\begin{figure}[h!]
\centering
\includegraphics [width=0.9 \columnwidth] {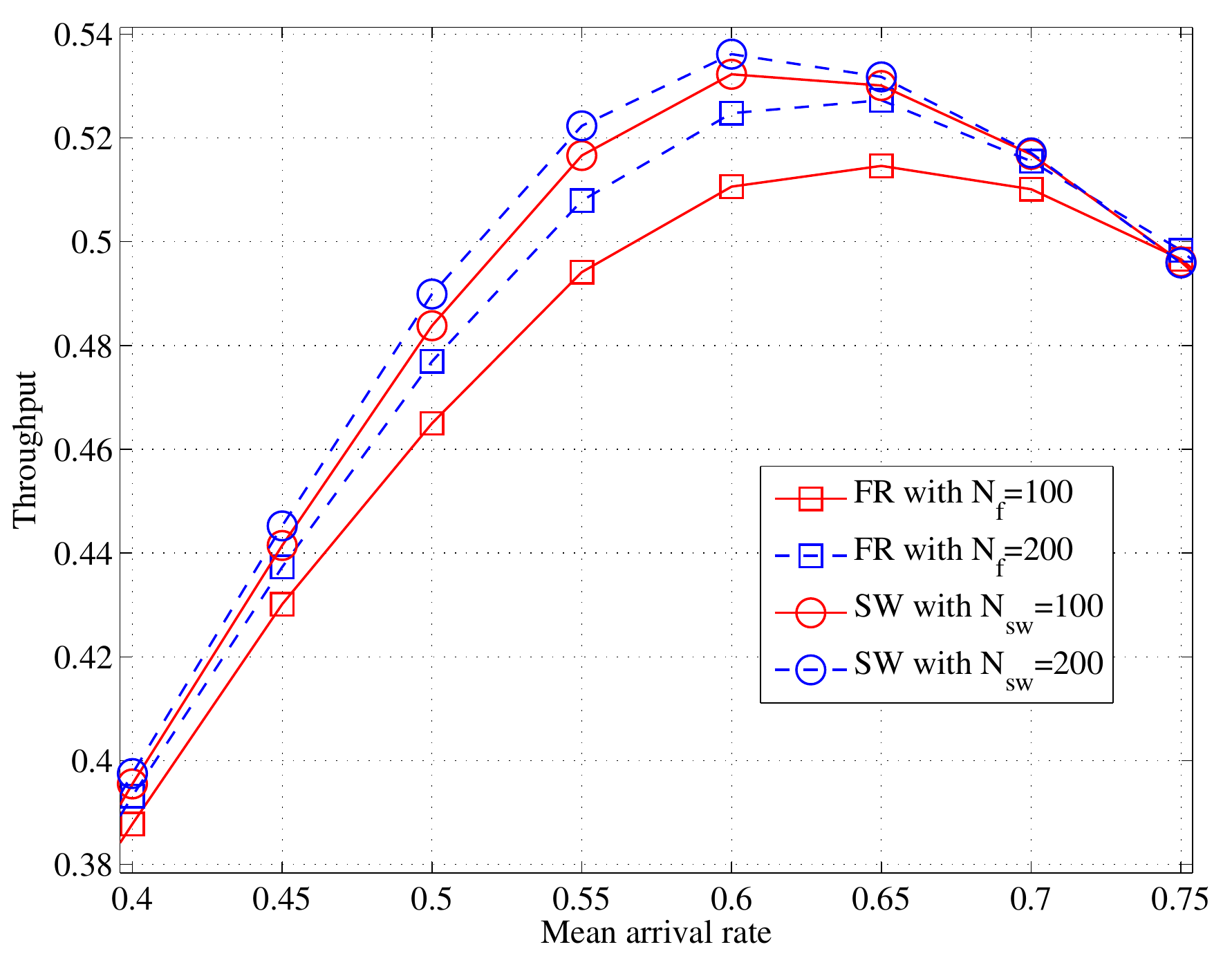}
\caption{Simulation results for the throughput in case of CRDSA (2 instances) with $N_{rx}=500$ slots for the SW case.}
\label{100200thrp}
\end{figure}

\begin{figure}[h!]
\centering
\includegraphics [width=0.9 \columnwidth] {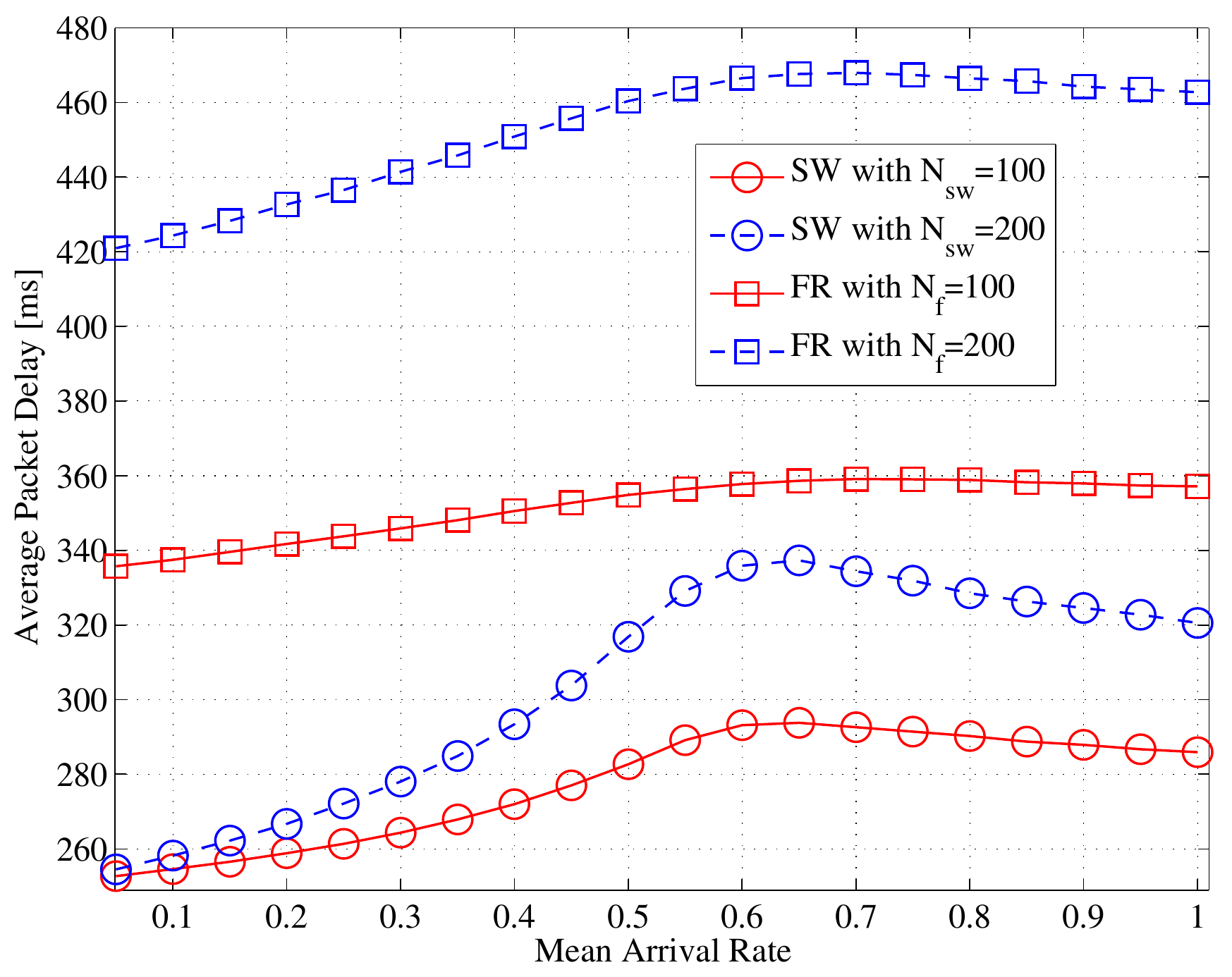}
\caption{Simulation results for the delay in case of CRDSA (2 instances) with $N_{rx}=500$ slots for the SW case.}
\label{100200delay}
\end{figure}

Also the choice of the number of slots for $N_f$ and $N_{sw}$ influences the resulting throughput, although with different dependences. Figure~\ref{100200thrp} shows that increasing the frame and sliding window size, FR's throughput is more influenced than SW's one.
Moreover Figure~\ref{100200delay} shows that while in the FR case the average packet delay has higher dependence on the frame size, the delay in case of SW is remarkably influenced by the value of $N_{sw}$ only for moderate to high arrival rates, but it is still less influenced by the size of $N_{sw}$ compared to FR.

\subsection{Packet Delay Distribution}

\begin{figure}[h!]
\centering
\includegraphics [width=1.01 \columnwidth] {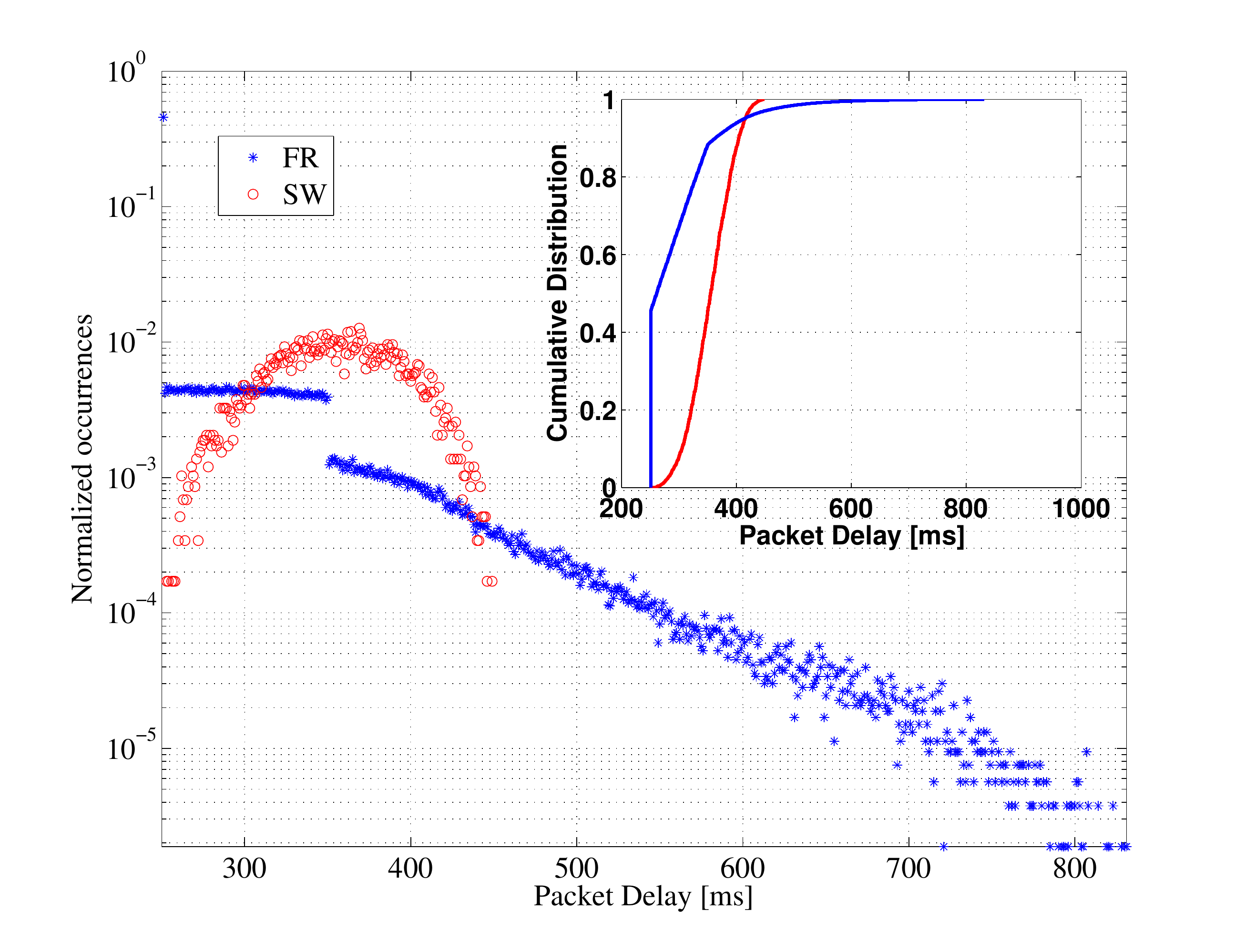}
\caption{Normalized packet delay occurrencies and cumulative distribution for correctly received packets in CRDSA with $N_{sw}=N_f=100$ slots, $\bar{\lambda}=0.6 [pkt/slot]$ and $N_{rx}=500$ slots for the SW case}
\label{delayDistr}
\end{figure}

Another important aspect in the comparison of FR and SW is the distribution of packet delay occurrencies. Figure~\ref{delayDistr} shows an example of packet delay occurrences normalized over the number of correctly received packets for SW and FR as well as its cumulative distribution, in order to verify the probability that a certain delay threshold is exceeded or not. The distribution of received packets for SW can be divided in three sets: 
\begin{itemize}
\item $T_p + T_{slot}$ (the most occurring value) represents the lowest achievable delay, i.e. the case in which the first instance is immediately decoded because it is alone in the slot or together with instances belonging to already decoded packets;
\item values of delay between $T_p$ + $2\ T_{slot}$ and $T_p + T_{slot}\cdot N_{sw}$ have more or less equal occurrence distribution, reflecting the fact that in this interval packet content is restored with more or less equally distributed probability among slots;
\item values greater than $T_p + T_{sw}$ present an exponential-like distribution highlighting that the number of occurrencies quickly decreases after $T_p + T_{slot}\cdot N_{sw}$.
\end{itemize}
Regarding FR, as expected the packet delay is distributed between  $T_p+T_{slot}$ and $T_p + 2\ T_f$. In particular, the distribution is almost symmetric, it has its maximum at $350 ms$ (that is $T_p+T_f)$ and an occurrence ratio that linearly decreases when considering values away from the maximum in both directions.
Consider now the cumulative probability distribution in the small window. Given a certain timeout $T_{to}$ for the packet delay, SW gets much better results than FR in terms of ratio of received packets with delay $\leq T_{to}$, up to timeout values close to $T_p + 2\ T_f$. However, it is necessary to verify that the better throughput performance of SW is not invalidated, considering the dependency shown in \ref{memsize}. 

\subsection{Overall Results}

\begin{figure}[t!]
\centering
\includegraphics [width=0.9 \columnwidth] {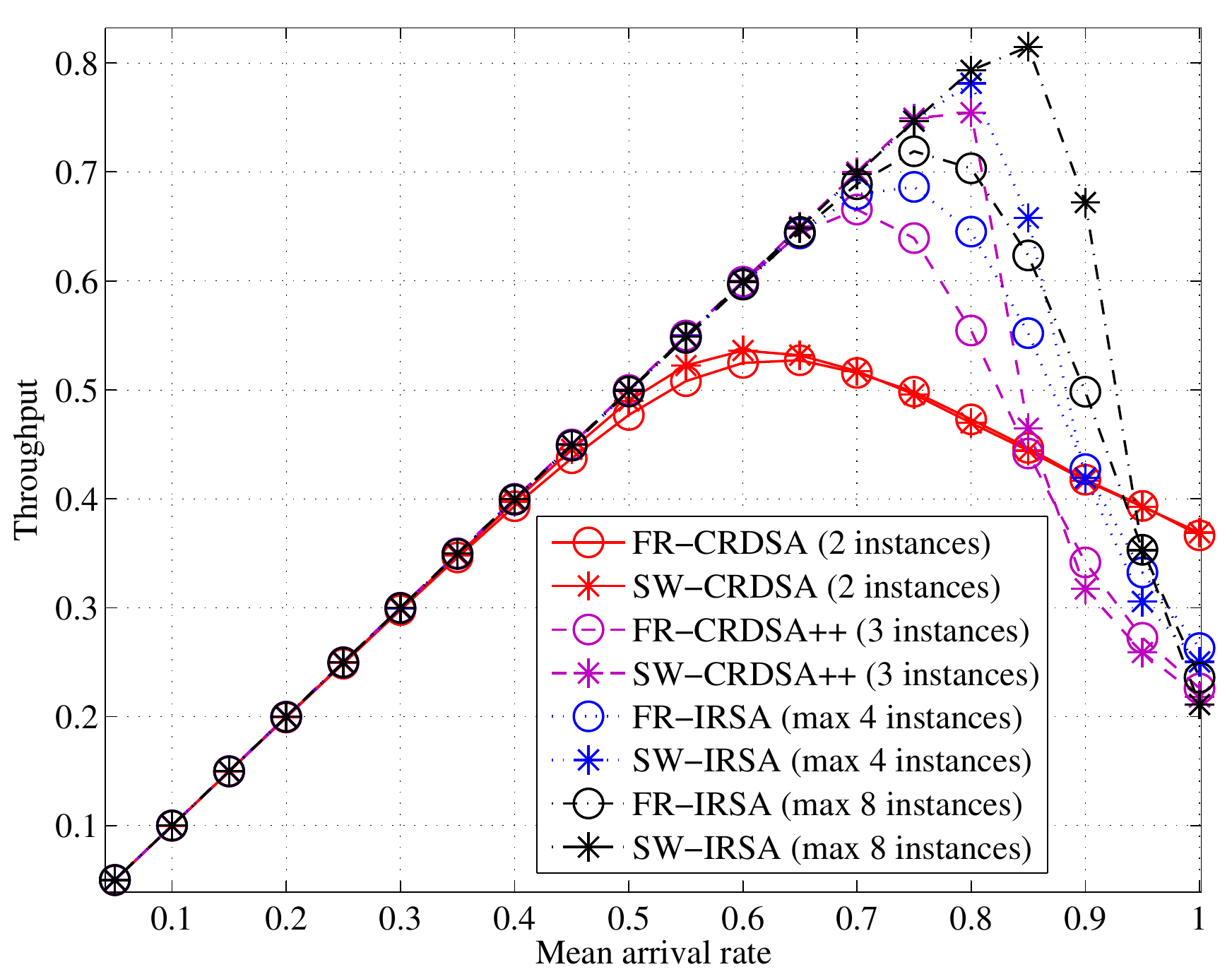}
\caption{Simulated throughput for FR and SW with  $N_f=N_{sw}=200$ slots and $N_{rx}=500$ slots for the SW case.}
\label{totFrSw}
\end{figure}

\begin{figure}[t!]
\centering
\includegraphics [width=0.9 \columnwidth] {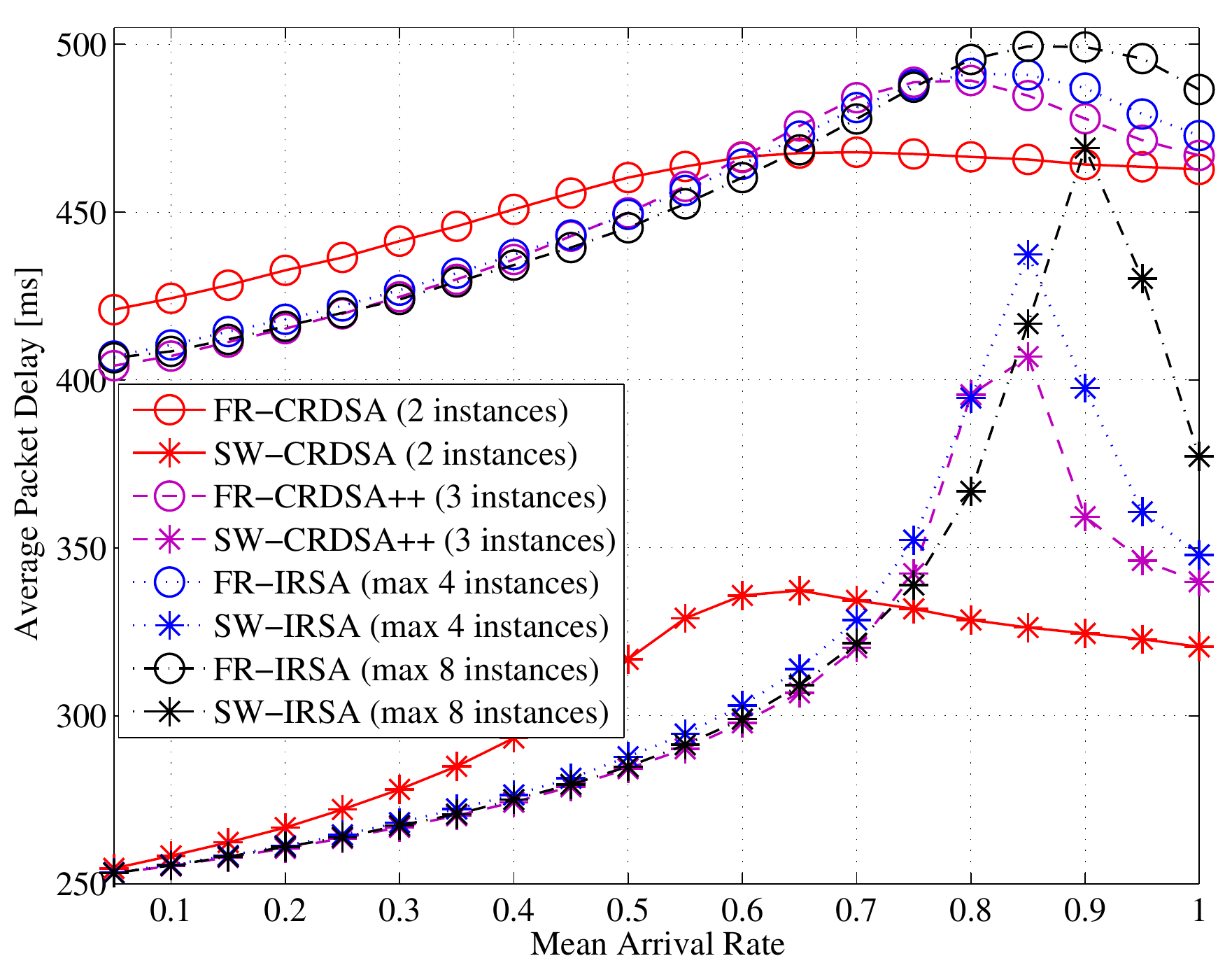}
\caption{Simulated delay for FR and SW with  $N_f=N_{sw}=200$ slots and $N_{rx}=500$ slots for the SW case.}
\label{totDelay}
\end{figure}

Finally let us compare the throughput and average packet delay performance of FR and SW for various regular and irregular burst repetition distributions. Figure~\ref{totFrSw} shows the throughput curves for different cases. As it can be seen, SW outperforms FR. In particular, while for SW-CRDSA a modest $2\%$ gain is obtained, for the other burst degree distributions a good $13\%$ gain is achieved. Moreover the peak throughput is shifted to bigger values of mean arrival rate (showing that SW operates at its optimal point in bigger traffic conditions than FR) and also the region in which the throughput is almost linear (corresponding to successful decoding of almost all packets sent) is extended to bigger mean arrival rates. The only part in which it would be convenient to use FR instead of SW is for very high mean arrival rates. However this is a zone of congestion in which it is not desirable to have the channel falling into because of the low throughput. Figure~\ref{totDelay} shows the average packet delay performance for the same cases considered in Figure~\ref{totFrSw}. As we can see, SW gets always much smaller delay than FR. In particular, the average packet delay for SW is at least $100\ ms$ less than the corresponding value for FR, up $\bar{\lambda} = 0.8$. Moreover it can be seen that the use of 2 instances per packet is the best burst degree choice when $\bar\lambda > 0.7$ . On the contrary, for smaller values a bigger mean number of instances pays off in terms of diminished average packet delay both in the FR and SW case.

\section{Conclusions and future work}

In this paper a new channel access scheme for Contention Resolution Diversity SA techniques have been introduced. This novelty shows a throughput performance up to 13\% greater than the one using frames; moreover, with this new access scheme, the throughput curve has a wider linear region and also the throughput peak is shifted to greater values of mean arrival rates. Also the average packet delay is greatly reduced, rendering this scheme attractive for delay critical applications especially in cases in which retransmission is not convenient or even impossible because of propagation delay issues as in satellite communications. The presented scheme can also be extended to the more general case in which also the first replica is distributed with equal probability over the sliding window set. However it is believed that this choice would not really improve the throughput while the delay performance would get worse due to additional time before the first replica is sent. 
Finally, we want to remark that presented simulations have been conducted on the MAC layer only, implicitly assuming same physical layer configuration and peak transmitting power for any burst degree distribution. However, as pointed out by one the anonymous reviewers, in satellite communication systems the performance can be limited by transmission power on board thus yielding to low SNR conditions and imperfect interference cancellation. Moreover, the satellite transponder has nonlinear characteristics that could as well disturb interference cancellation performance. For this reason, it is our aim to extend the content of this paper considering the constraint on the normalized efficiency (i.e. under average transmission power) as well as simulations implementing the physical layer with various SNR conditions and non-linear effects in order to evaluate the impact of imperfect interference cancellation.

\section*{Acknowledgments}
The authors would like to thank Gianluigi Liva for the useful discussions related to this work.

\end{document}